\documentclass[a4paper,fleqn,usenatbib,letters]{mnras}

\usepackage[dvipdfm]{graphicx,color}	
\usepackage{graphicx}	
\usepackage{amsmath}	
\usepackage{amssymb}	

\definecolor{mygray}{gray}{0.7}

\title[Quasar clustering in $\nu$GC]{Quasar clustering in a galaxy and quasar formation model based on ultra high-resolution $N$-body simulations}

\author[T. Oogi et al.]{
Taira Oogi,$^{1,2}$\thanks{E-mail: oogi@koshigaya.bunkyo.ac.jp (TO)}
Motohiro Enoki,$^{3}$
Tomoaki Ishiyama,$^{4}$
Masakazu A. R. Kobayashi,$^{5}$
\newauthor{ Ryu Makiya$^{6}$
and Masahiro Nagashima$^{1,2}$}
\\
$^{1}$Faculty of Education, Bunkyo University, 3337 Minami-Ogishima, Koshigaya-shi, Saitama 343-8511, Japan\\
$^{2}$Faculty of Education, Nagasaki University, 1-14, Bunkyo-machi, Nagasaki City, Nagasaki, 852-8521, Japan\\
$^{3}$Faculty of Business Administration, Tokyo Keizai University, 1-7-34, Minami-cho, Kokubunji-shi, Tokyo 185-8502, Japan\\
$^{4}$Institute of Management and Infomation Technologies, Chiba University, 1-33, Yayoi-cho, Inage-ku, Chiba, 263-8522, Japan\\
$^{5}$Research Center for Space and Cosmic Evolution, Ehime University, Bunkyo-cho, Matsuyama, Ehime, 790-8577, Japan\\
$^{6}$Institute of Astronomy, the University of Tokyo, 2-21-1, Osawa, Mitaka-shi, Tokyo, 181-0015, Japan\\
}

\date{Accepted XXX. Received YYY; in original form ZZZ}

\pubyear{2015}

\begin{document}
\label{firstpage}
\pagerange{\pageref{firstpage}--\pageref{lastpage}}
\maketitle

\begin{abstract}
We investigate clustering properties of quasars using a new version of our semi-analytic model of galaxy and quasar formation with state-of-the-art cosmological $N$-body simulations.
In this study, we assume that a major merger of galaxies triggers cold gas accretion on to a supermassive black hole and quasar activity.
Our model can reproduce the downsizing trend of the evolution of quasars.
We find that the median mass of quasar host dark matter haloes increases with cosmic time by an order of magnitude from $z=4$ (a few $10^{11} M_{\odot}$) to $z=1$ (a few $10^{12} M_{\odot}$), and depends only weakly on the quasar luminosity.
Deriving the quasar bias through the quasar--galaxy cross-correlation function in the model, we find that the quasar bias does not depend on the quasar luminosity, similar to observed trends.
This result reflects the fact that quasars with a fixed luminosity have various Eddington ratios and thus have various host halo masses that primarily determine the quasar bias.
We also show that the quasar bias increases with redshift, which is in qualitative agreement with observations.
Our bias value is lower than the observed values at high redshifts, implying that we need some mechanisms that make quasars inactive in low-mass haloes and/or that make them more active in high-mass haloes.
\end{abstract}




\section{Introduction}

Quasar clustering is one of the most important observational quantities to probe physical properties of quasars, providing information of the mass of dark matter (DM) haloes in which quasars reside.
This would be a strong constraints on quasar formation models \citep[e.g.,][]{2001ApJ...547...27H, 2001ApJ...547...12M}.
The clustering properties are quantified with the quasar bias, which is defined as the square root of the ratio of the two-point correlation function of the quasars, $\xi_\mathrm{Q}(r)$, to that of the DM, $\xi_{\mathrm{DM}}(r)$: $b_\mathrm{Q}(r) = \sqrt{\xi_\mathrm{Q} (r)/\xi_{\mathrm{DM}}(r)}$.
Comparing it with the bias of DM haloes by, for example, \citet{2001MNRAS.323....1S}, the typical DM halo mass of the quasars is derived.
The clustering properties have been reported using some large-scale surveys such as the Sloan Digital Sky Survey (SDSS; \citealt{2000AJ....120.1579Y}) and the 2dF QSO Redshift Survey (2QZ; \citealt{2000MNRAS.317.1014B}) 
(e.g., \citealt*{2004MNRAS.355.1010P}; \citealt{2005MNRAS.356..415C}; \citealt{2007ApJ...658...85M}; \citealt{2007AJ....133.2222S}; \citealt{2009MNRAS.397.1862P}; \citealt{2009ApJ...697.1634R}; \citealt{2015ApJ...803....4K}).
These studies suggest that, at low $z$ ($z\la2$), the typical DM halo mass of luminous quasars is $2-3\times 10^{12}~h^{-1}~M_{\odot}$, and does not depend on quasar luminosity.
At higher redshift ($z\ga3$), it is more massive than about $5\times10^{12}~h^{-1}~M_{\odot}$ \citep[][]{2007AJ....133.2222S, 2009ApJ...697.1656S}.
For more recent studies, \citet{2015ApJ...809..138I} and \citet{2015MNRAS.453.2779E} have investigated the bias of low-luminosity quasars and intermediate redshift ($2<z<3$) quasars.
Furthermore, at $z\sim4$--5, the overdensities of galaxies around quasars and radio galaxies have been discovered \citep[e.g.,][]{2006ApJ...640..574Z, 2008ApJ...673..143O}.

There are many works on correlations of the supermassive black hole (SMBH) mass with the stellar velocity dispersion of their host bulges \citep[e.g.,][]{2000ApJ...539L...9F, 2000ApJ...539L..13G, 2013ApJ...764..184M}, and with the bulge stellar mass (e.g., \citealt{1998AJ....115.2285M}, see also \citealt{2013ApJ...764..184M}).
These correlations should imply that the growth of SMBHs and their hosts are intimately linked.
To understand the correlations,
many semi-analytic models of galaxy formation including black hole mass growth and quasar activity have been proposed 
(e.g., \citealt{2000MNRAS.311..576K}; \citealt*{2003PASJ...55..133E}; \citealt{2005MNRAS.364..407C}; \citealt{2006MNRAS.373.1173F}; \citealt*{2007MNRAS.375.1189M}; \citealt{2008MNRAS.391..481S}).
Along this line, the quasar clustering should be investigated in the framework of the hierarchical formation of galaxies and quasars.

In previous theoretical studies on the the spatial clustering and environments of quasars in the context of hierarchical galaxy formation models 
(\citealt{2009MNRAS.396..423B}; \citealt*{2011MNRAS.413.1383D}; \citealt{2013MNRAS.436..315F}), the origin of the properties of the quasar clustering is not fully understood.
One of the main reasons is due to the low space density of quasars, which leads to large errors in the calculation of the clustering amplitudes.
For example, while the number density of galaxies brighter than the characteristic absolute magnitude is $\sim10^5~\mathrm{Gpc}^{-3}$ according to the rest-frame $V$-band luminosity function at $2.7<z<3.3$ (\citealt{2012ApJ...748..126M}),
that of quasars brighter than the break luminosity is $\sim100~\mathrm{Gpc}^{-3}$ according to the quasar luminosity function at $z=3$ (\citealt{2013ApJ...773...14R}).
Thus, the quasar number density is far smaller than the galaxy number density.
To overcome this situation, we need cosmological $N$-body simulations which have a high mass resolution and a large spatial volume for constructing DM halo merger trees.

In this Letter, we investigate the evolution of the large-scale quasar clustering using an updated-version of our semi-analytic model of galaxy and quasar formation, Numerical Galaxy Catalog ($\nu$GC) 
(\citealt{2005ApJ...634...26N}; \citealt{2014ApJ...794...69E}; see also \citealt{2015arXiv150807215M}, which is an updated version from the $\nu$GC).
We note that we construct DM halo merger trees using an ultralarge cosmological $N$-body simulation \citep{2015PASJ...67...61I} based on the $Planck$ cosmology.
This simulation has a substantially larger volume and a substantially higher mass resolution than the previous semi-analytic models, which allow us to investigate the quasar clustering with statistical significance.
In addition, our model naturally reproduces the downsizing trend of the quasar space density evolution \citep{2014ApJ...794...69E}.

The remainder of this Letter is organized as follows.
In Section~\ref{sec:model}, we outline our semi-analytic model and the methods of our analysis.
Section~\ref{sec:result} gives our predictions for the clustering of quasars in the range $0<z<5$.
In Section~\ref{sec:discussion}, we discuss the fueling mechanism of SMBHs comparing our results with observations, and summarize our results.

\section{Methods}
\label{sec:model}

\subsection{Numerical Galaxy Catalog}
\label{sec:nugc}

To investigate the quasar clustering, we use our semi-analytic model of galaxy and quasar formation, $\nu$GC \citep{2005ApJ...634...26N, 2003PASJ...55..133E, 2014ApJ...794...69E, 2015MNRAS.450L...6S}.
We construct merger trees of DM haloes using an ultralarge cosmological $N$-body simulation, $\nu^2$GC-M (\citealt{2015PASJ...67...61I}).
The simulation contains $4096^3$ DM particles in a comoving box of 560~$h^{-1}$~Mpc.
The mass of each particle is $2.20\times 10^8~h^{-1}~M_{\odot}$.
The adopted cosmological parameters are based on a $\Lambda$ cold dark matter cosmology, and are the ones obtained by the $Planck$ satellite \citep{2014A&A...571A..16P}: $\Omega_0=0.31,~\Omega_b = 0.048,~\lambda_0=0.69,~h=0.68,~n_s = 0.96$ and $\sigma_8 = 0.83$.
Further details of the simulation are given in \citet{2015PASJ...67...61I}.
$\nu$GC takes into account all the main processes involved in galaxy formation: (i) formation and evolution of DM haloes, (ii) radiative gas cooling and disc formation in DM haloes, (iii) star formation, supernova feedback and chemical enrichment, (iv) galaxy mergers.
Further details of our model of galaxy formation are given in \citet{2005ApJ...634...26N}.

Our model also includes formation and evolution of SMBHs and quasars.
Here, we briefly review our model.
Further details are given in \citet{2003PASJ...55..133E, 2014ApJ...794...69E}.

We assume that major mergers of galaxies trigger cold gas accretion on to SMBHs and quasar activity as well as starbursts \citep[e.g.,][]{2000MNRAS.311..576K, 2003PASJ...55..133E, 2005Natur.433..604D, 2008ApJS..175..356H}.
We define the major mergers as those mergers with mass ratios $M_2/M_1\geq0.2$, 
where $M_1$ and $M_2$ are the baryonic masses of the more and less massive galaxies, respectively.
During major mergers, we assume that SMBHs in progenitor galaxies merge instantaneously and that a fraction of the cold gas, which is proportional to the total mass of stars newly formed during the starburst, is accreted on to the merged SMBH.
Under this assumption, the cold gas mass accreted on the SMBH is given by
\begin{equation}
    M_{\mathrm{acc}} = f_{\mathrm{BH}} \Delta M_{*,\mathrm{burst}},
\end{equation}
where $\Delta M_{*,\mathrm{burst}}$ is the total stellar mass formed during the starburst and $f_{\mathrm{BH}} = 0.0067$ is a parameter chosen to match the observed relation between masses of host bulges and SMBHs at $z=0$ found by \citet{2013ApJ...764..184M}.

The cold gas accretion on to an SMBH during a starburst leads to quasar activity.
We assume that a fixed fraction of the rest mass energy of the accreted gas is radiated in the $B$-band
and that the evolution of quasar $B$-band luminosity follows the light curve as a function of the elapsed time from the major merger as follows:
\begin{equation}
    L_B (t) = \frac{\epsilon_B M_{\mathrm{acc}} c^2}{t_{\mathrm{life}}} \exp(-t/t_{\mathrm{life}}),
	\label{eq:lightcurve}
\end{equation}
where $\epsilon_B$ is the radiative efficiency in the $B$-band, $t_{\mathrm{life}}$ is the quasar lifetime, and $c$ is the speed of light.
We assume that $t_{\mathrm{life}}$ scales with the dynamical time-scale of the host DM halo.
We choose two parameters, $\epsilon_B$ and the present quasar lifetime $t_{\mathrm{life}}(z=0)$, to match the estimated luminosity function in our model with the observed $B$-band luminosity function of quasars at $z=2$.
We obtain $\epsilon_B=0.0033$ and $t_{\mathrm{life}}(z=0)=1.5\times10^7$ yr
\footnote{These values are slightly different from those of \citet{2014ApJ...794...69E}, while the prescriptions are the identical
 with each other.
This is due to the changes of the cosmology from the $WMAP7$ to $Planck$ and of the criterion for major mergers of galaxies,
the latter of which is found to be dominant.
These changes do not affect the results in \citet{2014ApJ...794...69E} qualitatively.}.

Our model can reproduce the downsizing trend of the quasar evolution \citep{2014ApJ...794...69E}.
This outcome allows us to prove the quasar clustering with our model.

\subsection{Clustering analysis}
\label{sec:clustering_analysis}

Here, we describe the method to derive the quasar bias, $b_\mathrm{Q} = \sqrt{\xi_\mathrm{Q}/\xi_{\mathrm{DM}}}$, with cross-correlation functions.
Since the space density of quasars is too low for a statistically significant autocorrelation study, we adopt a cross-correlation analysis instead.
According to \citet{2007ApJ...654..115C}, we calculate the quasar-galaxy cross-correlation function, $\xi_{\mathrm{QG}}$, using our samples of quasars and galaxies as follows:
\begin{equation}
    \xi_{\mathrm{QG}} (r) = \frac{\mathrm{QG} (r)}{\mathrm{QR} (r)} - 1,
\end{equation}
where $\mathrm{QG}(r)$ and $\mathrm{QR}(r)$ are quasar--galaxy and quasar--random pairs at a given separation, $r$, respectively.
These pair counts are normalized by $n_\mathrm{G}$ and $n_\mathrm{R}$ which are the mean number densities in the full galaxy and random catalogues, respectively.

To measure the quasar bias, we also calculate the galaxy autocorrelation function $\xi_\mathrm{G}$ using our sample of galaxies with $B$-band magnitudes $M_{B} - 5\log h < -20.0$, as follows:
\begin{equation}
    \xi_\mathrm{G} (r) = \frac{\mathrm{GG}(r)}{\mathrm{GR}(r)} - 1,
\end{equation}
where $\mathrm{GG}(r)$ and $\mathrm{GR}(r)$ are galaxy--galaxy and galaxy--random pairs at $r$ \citep{1983ApJ...267..465D}, respectively.
These pair counts are also normalized by $n_\mathrm{G}$ and $n_\mathrm{R}$, respectively.

Then, we estimate the quasar bias, $b_{\mathrm{Q}}$, from $\xi_{\mathrm{QG}} (r)$ and $\xi_\mathrm{G} (r)$.
Assuming a linear bias, we calculate the quasar bias as follows:
\begin{equation}
    b_{\mathrm{Q}} (r) = \frac{\xi_{\mathrm{QG}}(r)}{\sqrt{\xi_\mathrm{G}(r)\xi_{\mathrm{DM}}(r)}},
\end{equation}
where $\xi_{\mathrm{DM}}(r)$ is the autocorrelation function for the DM, which is calculated with our cosmological $N$-body simulations.

To compare our bias with observations, we have corrected the bias factors $b$ observationally estimated under different cosmologies to those under the $Planck$ cosmology we adopt using the following equation:
\begin{equation}
    b' (z) = \frac{\sigma_8 D(z)}{\sigma'_8 D'(z)} b,
\end{equation}
where $b' (z)$ is the bias factor after correction, $\sigma_8$ and the growth factor $D(z)$ are those for each observation, and $\sigma'_8$ and $D'(z)$ are those for the $Planck$ cosmology.

\section{Results}
\label{sec:result}

\subsection{Quasar host halo mass}
\label{subsec:halo_mass}

\begin{figure}
	\centering
	\includegraphics[width=0.9\columnwidth]{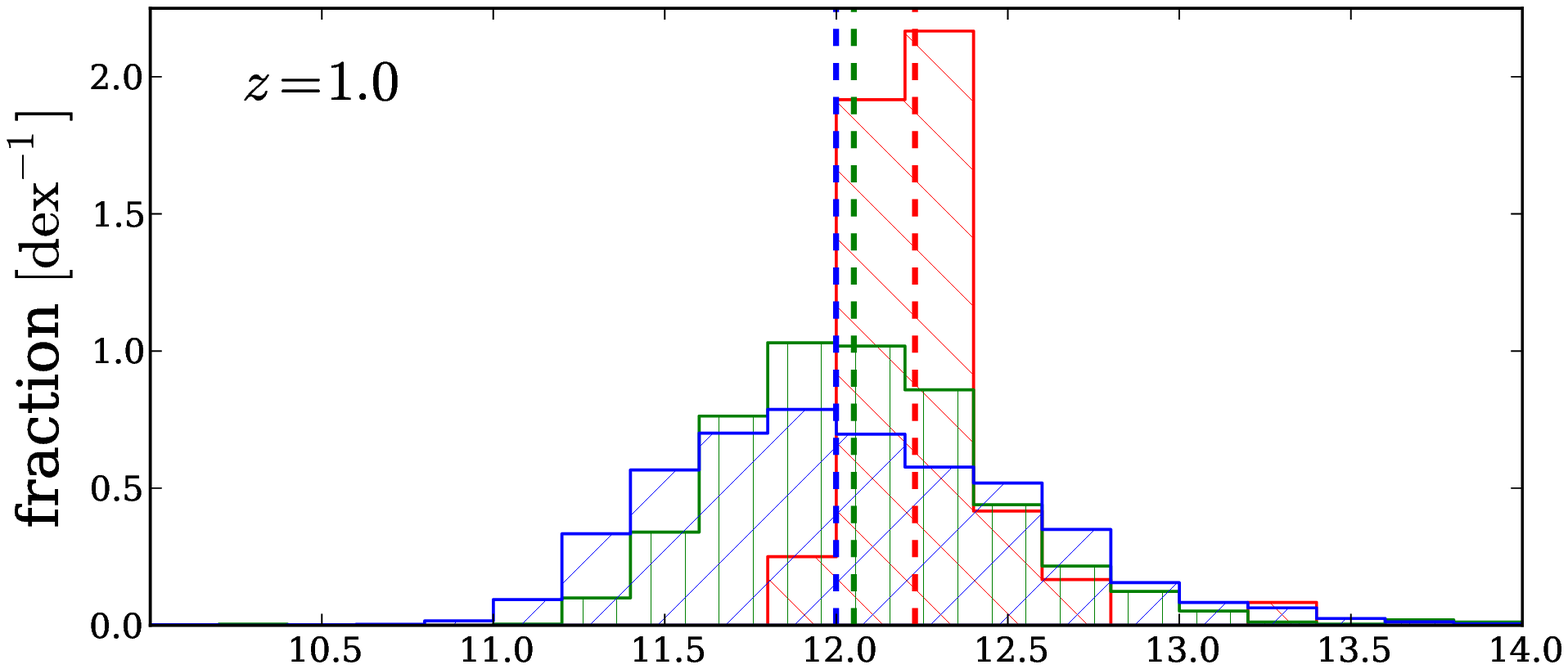}\\
	\includegraphics[width=0.9\columnwidth]{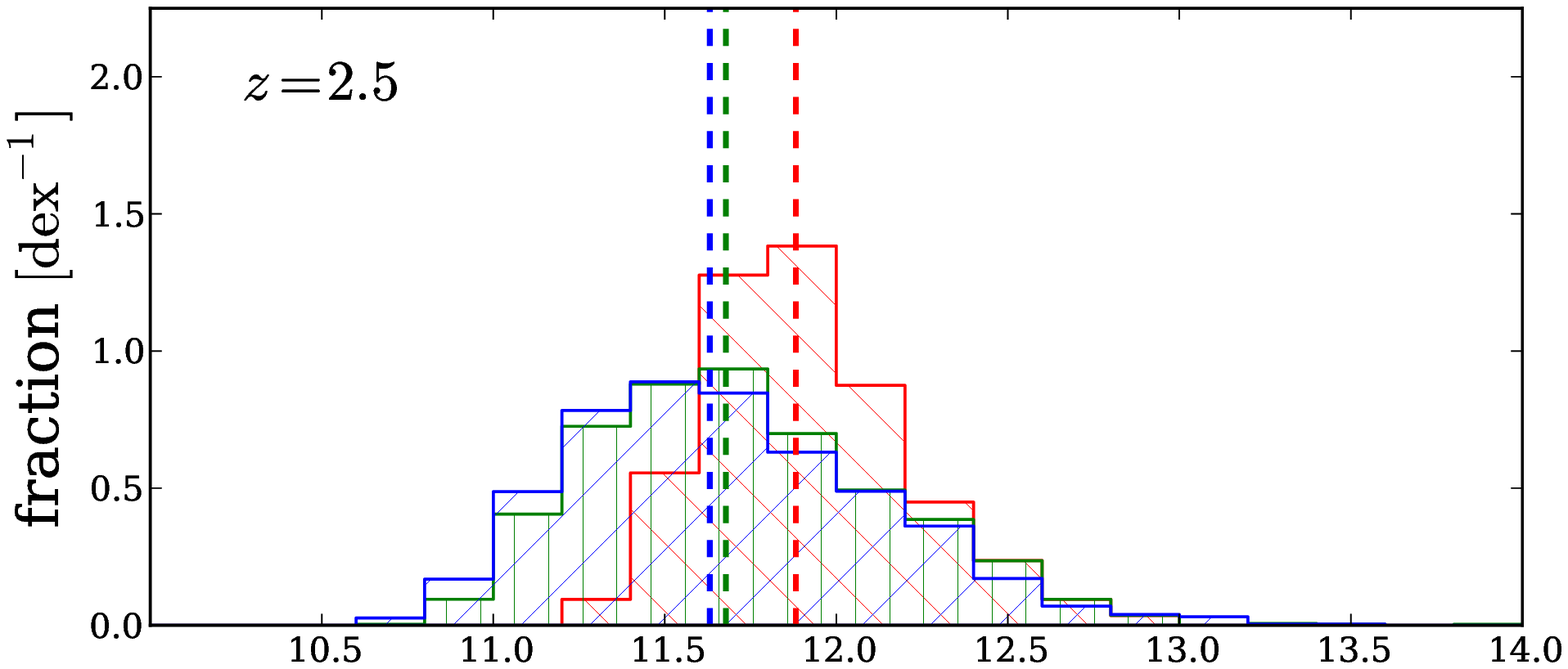}\\
	\hspace{0.0mm}
	\includegraphics[width=0.9\columnwidth]{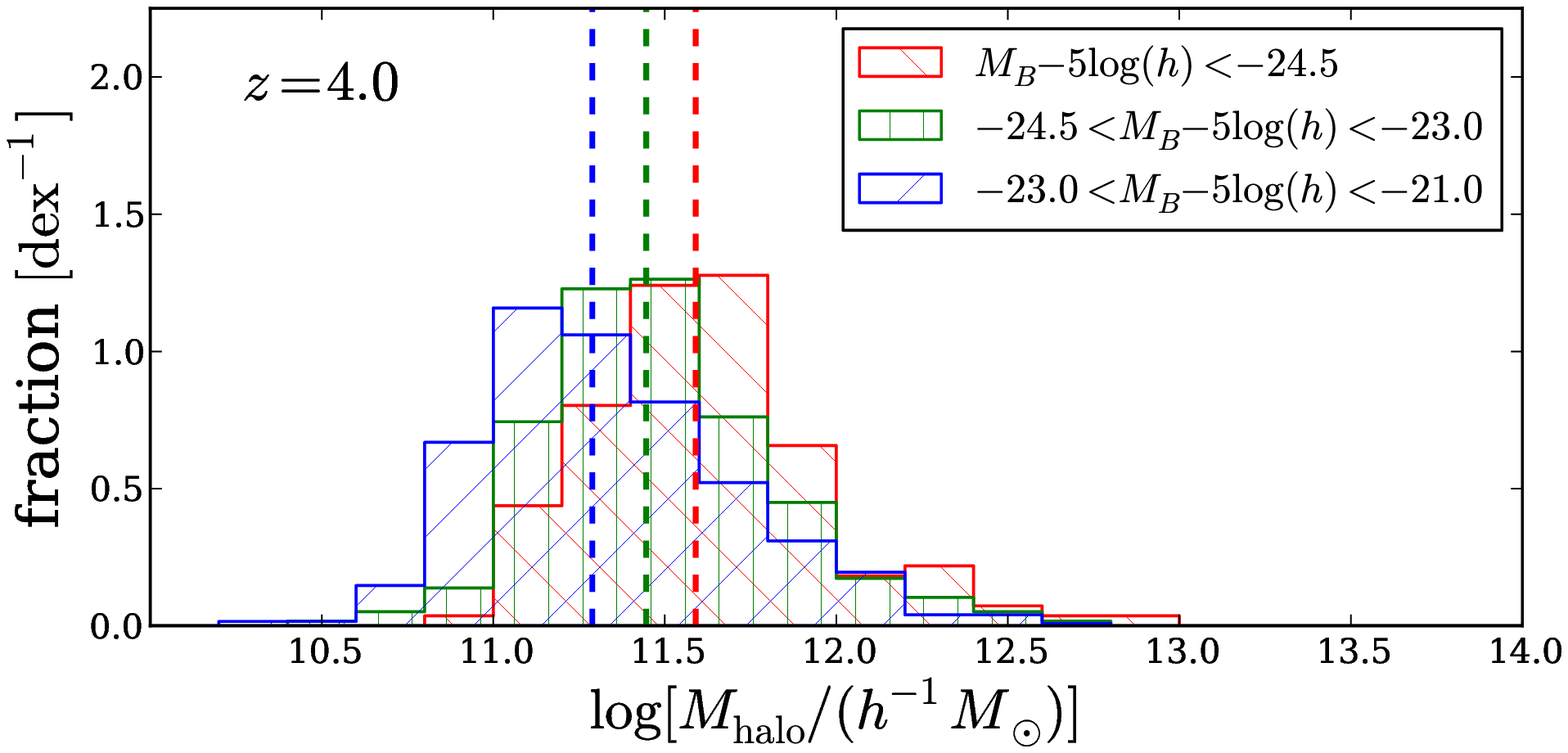}
    \caption{
    Mass distributions of DM haloes hosting bright (red), intermediate (green) and faint (blue) quasars at $z=1.0$ (top), $2.5$ (middle), and $4.0$ (bottom).
    The vertical dashed lines denote the median masses of the distributions.}
    \label{fig:m_halo_dist}
\end{figure}

The host halo mass of quasars is one of the most fundamental properties characterizing the environments of quasars.
Because our model is based on the merger trees of DM haloes, we directly derive the mass distribution of the quasar host haloes.
Fig.~\ref{fig:m_halo_dist} shows the mass distributions of the quasar host haloes for three quasar-luminosity bins and their median halo masses (dashed lines).
We define bright, intermediate, and faint quasars as those with $B$-band magnitudes $M_{B} - 5\log h < -24.5$, $-24.5 < M_{B} - 5\log h < -23.0$ and $-23.0 < M_{B} - 5\log h < -21.0$, respectively.
Note that we have confirmed that there is little difference between the median and the mean mass.
Each mass distribution has a peak around its median.
This indicates that the quasar host haloes have a characteristic halo mass which is represented by the median or mean halo mass.
The characteristic mass increases with cosmic time by an order of magnitude (from a few $10^{11} M_{\odot}$ at $z=4$ to a few $10^{12} M_{\odot}$ at $z=1$).
This result is consistent with other theoretical studies which are based on the hierarchical structure formation \citep[][]{2011MNRAS.413.1383D, 2013MNRAS.436..315F}.
In addition, the characteristic mass does not depend significantly on quasar luminosity.
As the characteristic halo mass primarily determines the quasar bias, we analyse the quasar bias in the next section.

\subsection{Quasar bias}

\begin{figure}
	\centering
	\includegraphics[width=0.95\columnwidth]{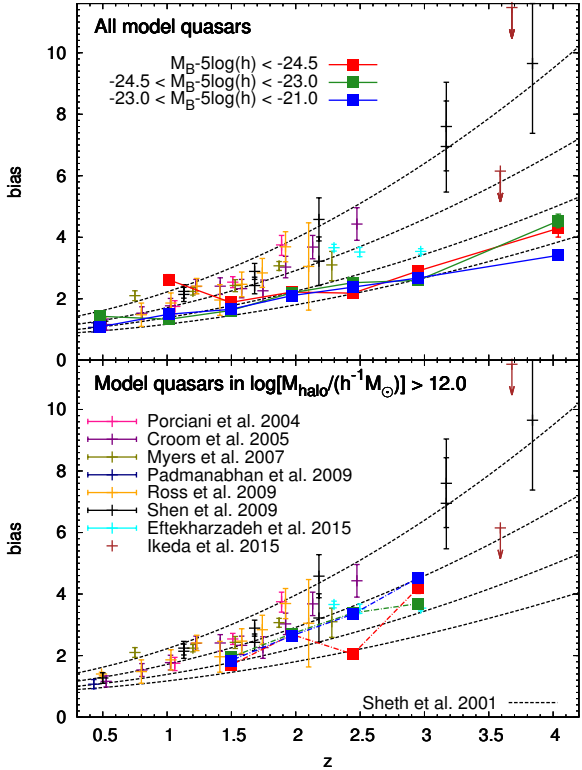}
    \caption{
    Top: redshift evolution of the bias of bright (red), intermediate (green) and faint (blue) quasars.
    The dashed lines are halo bias factor evolution for fixed halo mass of $\log [M_{\mathrm{halo}}/(h^{-1}~M_{\odot})] = 11.5$, 12.0, 12.5 and 13.0 from bottom to top, respectively, using the \citet{2001MNRAS.323....1S} fitting formula with the $Planck$ cosmology.
    Observational results are also plotted (plus signs and error bars).
    Bottom: the same as the top panel, but for only model quasars which are hosted DM haloes ($M_{\mathrm{halo}}>10^{12}~h^{-1}~M_{\odot}$).
    }
    \label{fig:z_bias}
\end{figure}

We then present the properties of the quasar bias calculated using the method we described in Section~\ref{sec:clustering_analysis}, and compare them with observations.
In the top panel of Fig.~\ref{fig:z_bias}, we show the redshift evolution of the quasar bias of bright, intermediate, and faint quasars.
As shown in this figure, the quasar bias increases with redshift from $b\sim1$ at $z=1$ to $b\sim4$ at $z=4$.
This increase with redshift agrees with that of the bias of observed Lyman break galaxies in similar redshift ranges (\citealt{2004ApJ...611..685O}).
We have also confirmed that the bias of our sample galaxies with the $B$-band magnitude $M_{B} - 5\log h < -20.0$ agrees with the observation.
On the other hand, it has only weak dependence on quasar luminosity.
This weak luminosity-dependent clustering is in agreement with previous theoretical works \citep[][]{2006ApJ...641...41L, 2007ApJ...662..110H, 2008ApJS..175..356H, 2009MNRAS.394.1109C, 2009ApJ...704...89S, 2013ApJ...762...70C}.
In our model, this weak dependence reflects the weak luminosity dependence of the host halo mass as shown in Section~\ref{subsec:halo_mass}.
The halo mass estimated from the quasar bias using the formula of \citet{2001MNRAS.323....1S} is in agreement with the characteristic halo mass directly from our model in Section~\ref{subsec:halo_mass}.
This means that the quasar bias is determined primarily by the halo bias of the quasar host halo with the characteristic mass.
Moreover, this result supports the observational procedure for estimating the mass of the DM haloes which host quasars with the quasar bias.

In Fig.~\ref{fig:z_bias}, we also plot some observational results for comparison.
Almost all observational results used quasars obtained by the SDSS and the 2QZ.
For the SDSS DR 5 sample (\citealt{2009ApJ...697.1634R}; \citealt{2009ApJ...697.1656S}), the mean magnitude is $M_B-5\log h \approx -25.5$ at $z\sim2$.
Here, we convert the $K$-corrected absolute $i$-band magnitude $M_{i}(z=2)$ to $M_B$ using the conversions given in \citet{2005MNRAS.356..415C} and \citet{2006AJ....131.2766R}.
For the 2QZ sample (\citealt{2004MNRAS.355.1010P}; \citealt{2005MNRAS.356..415C}), the mean magnitude is $M_B-5\log h \approx -24.5$ at $z\sim2$ using the conversion from the $b_J$ band magnitude to $M_B$.
These magnitudes are comparable to those of our sample.
While the 2QZ sample has lower mean magnitude than the SDSS sample, the bias measurements of these two samples are similar.
This indicates that the quasar bias does not depend significantly on quasar luminosity.

The quasar bias by our model is in agreement with the observations at low $z$ ($z\la1.5$).
In addition, the weak luminosity-dependent clustering is also in agreement with the observations.
At high $z$, however, our results cannot reproduce the observed clustering of quasars, in particular $z>3$.
This would be caused by the disagreement in the typical DM halo mass, for example, $\sim 3\times 10^{12}~h^{-1}~M_{\odot}$ at $2\la z \la 2.5$, and $\sim 10^{13}~h^{-1}~M_{\odot}$ at $3 \la z$ in the observations.
We discuss this disagreement of our model in Section~\ref{sec:discussion}.
We note that, in the $Planck$ cosmology, the DM halo mass estimated from the observed quasar bias tends to be more massive than that originally estimated in each observational study.

\subsection{DM halo mass, quasar magnitude and Eddington ratio}

\begin{figure}
	 \centering
	\includegraphics[width=0.9\columnwidth]{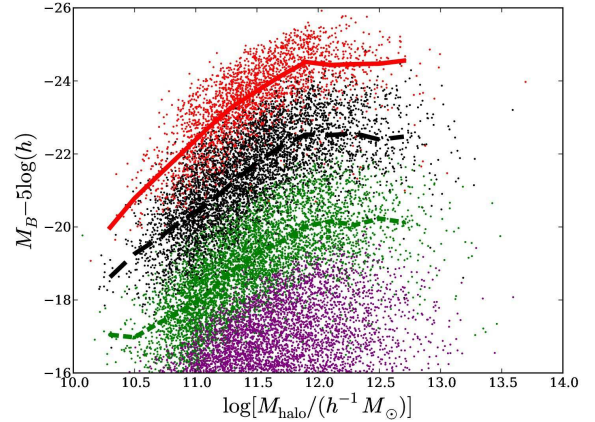}
    \caption{Relation between the DM halo mass and quasar $B$-band magnitude in our quasar sample at $z=2.5$.
    Red, black, green and purple dots denote the quasars with $B$-band Eddington ratios, $f_{\mathrm{Edd},B} \equiv (L_B / L_{\mathrm{Edd}})$, $-1 < \log f_{\mathrm{Edd},B}$, $-2 < \log f_{\mathrm{Edd},B} < -1$, $-3 < \log f_{\mathrm{Edd},B} < -2$ and $\log f_{\mathrm{Edd},B} < -3$, respectively.
    The solid, long-dashed and short-dashed lines show the medians of the $M_{\mathrm{halo}}-M_{B}$ correlation for quasars with $-1 < \log f_{\mathrm{Edd},B}$, $-2 < \log f_{\mathrm{Edd},B} < -1$, and $-3 < \log f_{\mathrm{Edd},B} < -2$, respectively.
    (A colour version of this figure is available in the online journal.)
    }
    \label{fig:m_halo_edd_z2p5}
\end{figure}

Our results show that the mass of DM haloes hosting quasars and the quasar bias do not depend significantly on quasar luminosity.
To clarify this origin, we examine the relation among the DM halo mass, quasar magnitude, and $B$-band Eddington ratio, $f_{\mathrm{Edd},B} \equiv (L_B / L_{\mathrm{Edd}})$.
In Fig.~\ref{fig:m_halo_edd_z2p5}, we plot the DM halo mass and $M_B$ relation for our quasar sample at $z=2.5$.
We divide the sample into four $f_{\mathrm{Edd},B}$ bins, and plot the subsamples with different colours.
As with in Fig.~\ref{fig:m_halo_dist}, there is no definite correlation between the DM halo mass and $M_B$.
On the other hand, we find the power law relationship between the quasar $B$-band luminosity and the DM halo mass at fixed $f_{\mathrm{Edd},B}$.
The quasars in the largest $f_{\mathrm{Edd},B}$ bin are close to their peak luminosity, while those in the smaller $f_{\mathrm{Edd},B}$ bins are in the decaying phase where quasars decrease their luminosities to zero.
Thus, the quasars evolve downward with time in Fig.~\ref{fig:m_halo_edd_z2p5}.
Consequently, quasars having various magnitudes reside in DM haloes with similar masses.
For this reason, the DM halo mass does not correlate with quasar magnitude.
This conclusion is consistent with the model of \citet{2006ApJ...641...41L}.

Although one may think that quasars at $z>3$ taken from flux-limited samples of \citet{2007AJ....133.2222S, 2009ApJ...697.1656S}
reside in more massive DM haloes because of their very high luminosities, our model shows that 
the quasar host halo mass does not depend significantly on quasar luminosity even at high-$z$ and at the 
luminous end. Therefore, it is not likely that the strong quasar bias at $z>3$ can be explained simply 
by their high luminosities.

We note that in our model, quasars have various Eddington ratios and luminosities due to a variety of elapsed times from the beginning of the quasar activity.
However, the details of the accretion flow also cause the difference of the quasar luminosity \citep*[e.g.,][]{2015MNRAS.452.1922P}.
We should take into account this possibility in future studies.

\section{Summary and Discussion}
\label{sec:discussion}

In this Letter, we have investigated the large-scale clustering of quasars, including its redshift and luminosity dependence using our semi-analytic model, $\nu$GC.
We have shown that the quasar bias has no significant dependence on quasar luminosity and increases with redshift.
We have also found that the quasar bias is primarily determined by the halo bias of the quasar host halo.

We have found that the model bias underpredicts the observed ones at $z\ga2$.
The underprediction is found to become smaller if we remove the model quasars within host haloes less than $10^{12}~h^{-1}~M_{\odot}$ as shown in the bottom panel of Fig.~\ref{fig:z_bias}.
This suggests that there are some mechanisms that inhibit quasar formation within low-mass haloes and/or that enhance the quasar activity within high-mass haloes.

Comparing the previous semi-analytic models \citep{2009MNRAS.396..423B, 2013MNRAS.436..315F}, the DM halo merger trees we use are based on a cosmological $N$-body simulation which has a larger volume and a higher mass resolution.
Furthermore, we can derive the quasar bias more accurately using the quasar--galaxy cross-correlation function, since investigating the quasar clustering by measuring the cross-correlation function can provide more accurate measurements of clustering amplitude of quasars, as the space density of galaxies is much higher.
In contrast, \citet{2009MNRAS.396..423B} investigated the quasar clustering using the quasar bias derived from the quasar--quasar autocorrelation function.
However, their predicted quasar bias has large errors because the errors on the clustering amplitudes estimated from the quasar autocorrelation function are large due to their low space density.
\citet{2013MNRAS.436..315F} also investigated the evolution of the mass of quasar host haloes and the quasar bias, although they did not directly use the correlation function to derive the quasar bias.
Their redshift evolution of DM halo mass is consistent with our model.

Our results presented here suggest that the triggering mechanism of quasars at high $z$ may be different from that at low $z$.
Previous theoretical studies have also pointed out the strong clustering of quasars at high $z$ as a challenging problem.
Using a simple model which is constrained by the clustering and abundance of quasars, \citet*{2008MNRAS.390.1179W} claimed  that the strong clustering measured at $z\sim4$ is difficult to understand 
unless quasar duty cycles are high and intrinsic scatter in the relation between the quasar luminosity and the halo mass is small (see also \citealt{2010ApJ...718..231S}).
Further observations of high-redshift and low-luminosity quasars with Hyper Suprime-Cam \citep[][]{2006SPIE.6269E..0BM, 2012SPIE.8446E..0ZM} will allow us to make accurate comparison to the model we introduced here and to estimate the role of major mergers as a triggering process of quasars.

\section*{Acknowledgements}

We are grateful to the referee for providing constructive comments.
We thank M. Akiyama, H. Ikeda, T. Nagao, N. Kashikawa, N. Kawakatu, T. Okumura, Y. Matsuoka, T. Okamoto and K. Wada for useful comments and discussion.
This study has been funded by MEXT/JSPS KAKENHI Grant Number 25287041, 15K12031 and by Yamada Science Foundation, MEXT HPCI STRATEGIC PROGRAM.
RM has been supported by the Grant-in-Aid for JSPS Fellows.




\bibliographystyle{mnras}
\bibliography{ref} 








\bsp	
\label{lastpage}
\end{document}